\documentclass[conference]{IEEEtran}

\usepackage{tabularx,booktabs}
\newcolumntype{C}{>{\centering\arraybackslash}X} 
\usepackage{lipsum} 

\usepackage{cite}
\usepackage{amsmath,amssymb,amsfonts}
\usepackage{algorithmic}
\usepackage{textcomp}
\usepackage{xcolor}
\usepackage{amsmath}
\usepackage{enumitem}

\usepackage[pdftex]{graphicx}
\DeclareGraphicsExtensions{.pdf,.jpeg,.png}
\usepackage{float}
\usepackage[outdir=Figures/]{epstopdf}
\usepackage{balance}
\usepackage{flushend}
\usepackage{etoolbox}

\usepackage{algorithm}

\usepackage{url}
\usepackage{hyperref}

\def\BibTeX{{\rm B\kern-.05em{\sc i\kern-.025em b}\kern-.08em
    T\kern-.1667em\lower.7ex\hbox{E}\kern-.125emX}}

\def\isarxiv{}

\ifdefined\isarxiv
\makeatletter
\def\ps@IEEEtitlepagestyle{%
  \def\@oddfoot{%
    \hbox to \textwidth{%
      \hfil
      \parbox{\textwidth}{\centering\scriptsize
      \textcopyright~2025 IEEE. Personal use of this material is permitted. Permission from IEEE must be obtained for all other uses, in any current or future media, including reprinting/republishing this material for advertising or promotional purposes, creating new collective works, for resale or redistribution to servers or lists, or reuse of any copyrighted component of this work in other works. DOI: \href{https://doi.org/10.1109/MOCAST65744.2025.11083895}{10.1109/MOCAST65744.2025.11083895}}%
      \hfil
    }%
  }%
  \def\@evenfoot{}%
}
\makeatother
\fi

\begin{document}

\ifdefined\isarxiv
\title{Efficient Deployment of CNN Models on Multiple In-Memory Computing  Units}
\else
\title{Efficient Deployment of CNN Models on multiple In-Memory Computing  Units}
\fi

\author{\IEEEauthorblockN{Eleni Bougioukou and Theodore Antonakopoulos}
\IEEEauthorblockA{\textit{Department of Electrical and Computer Engineering}\\ 
\textit{University of Patras, Greece} \\
\mbox{bougioukou@upatras.gr, antonako@upatras.gr}
}
}



\maketitle

\begin{abstract}
 In-Memory Computing (IMC) represents a paradigm shift in deep learning acceleration by mitigating data movement bottlenecks and leveraging the inherent parallelism of memory-based computations. The efficient deployment of Convolutional Neural Networks (CNNs) on IMC-based hardware necessitates the use of advanced task allocation strategies for achieving maximum computational efficiency. In this work, we exploit an IMC Emulator (IMCE) with multiple Processing Units (PUs) for investigating how the deployment of a CNN model in a multi-processing system affects its performance, in terms of  processing rate and latency. For that purpose, we introduce the Load-Balance-Longest-Path (LBLP) algorithm,  that dynamically assigns all CNN nodes to the available IMCE PUs, for maximizing the processing rate and minimizing latency  due to efficient resources utilization. We are benchmarking LBLP against other alternative scheduling strategies for a number of CNN models and experimental results demonstrate the effectiveness of the proposed algorithm.
\end{abstract}


\section{Introduction}
With the rapid growth of the Internet of Things (IoT) and Cloud Computing, there is a growing need for efficient deep learning models that can operate on diverse computing platforms, ranging from resource-constrained edge devices to high-performance data centers. Among others, Convolutional Neural Networks (CNNs) have become a cornerstone of deep learning\cite{li2020surveyconvolutionalneuralnetworks}, driving advances in image classification, object detection, and other computer vision tasks. The computational complexity of CNNs, especially for large-scale models, has necessitated the development of specialized hardware accelerators. Devices such as Neural Processing Units (NPUs) on microcontrollers \cite{stm32n6}, \cite{intel_movidius_myriad}, Digital Processing Units (DPUs) in Field-Programmable Gate Arrays (FPGAs) \cite{xilinx_alveo}, \cite{xilinx_versal},  Graphics Processing Units (GPUs) \cite{nvidia_a100} and Tensor Processing Units (TPUs) \cite{google_edge_tpu} have been developed to handle the computationally intensive tasks of CNNs, mainly convolutions and matrix multiplications. 

The digital accelerators face significant challenges related to data movement and energy consumption.  In-Memory Computing (IMC) \cite{analog_or_digital} has emerged as a promising alternative, addressing these challenges by performing computations directly within memory arrays. IMC drastically reduces data transfers, leading to improved throughput and higher energy efficiency. Many IMC devices use technology that performs the computations in the analog domain (AIMC) \cite{bowen2023analoginmemorycomputearchitectures}, with the advantage of energy efficiency and massive parallelization but with the presence of analog noise. An alternative approach for avoiding the hurdles of AIMC is to use digital IMC (DIMC) \cite{Hybrid_AIMC_DIMC}, with noise-free computation but with less parallelization. Currently hybrid devices with multiple IMCs and DPUs have been proposed, i.e. in  \cite{neurosoc}.

While using IMC results to improved efficiency, the effective scheduling of CNN workloads remains a challenge. Task mapping, often referred to as node mapping in neural networks, directly impacts latency, processing rate, and in some cases power dissipation. Effective scheduling strategies must balance computation and communication overhead to optimize execution.  In this work we address this research topic for CNNs.

Section \ref{CNNm2Inf} discusses the mapping of CNN models to inference engines resources, while Section \ref{IMCE_architecture} describes the IMCE platform we used for experimentation architecture. Section \ref{Algorithms} presents the Load Balancing Longest Path (LBLP) algorithm, along with three additional scheduling algorithms. All these algorithms are evaluated in Section \ref{PerfResults}, using ResNet and YOLO networks.

\section{From CNN Models to CNN Inference} \label{CNNm2Inf}

Deploying CNNs for real-time inference requires efficient allocation strategies that map computational nodes onto the available inference engines, named also as processing units (PUs). The challenge lies in optimally distributing the model nodes/tasks across the available PUs to maximize throughput and minimize latency. 

Several task mapping algorithms have been proposed for heterogeneous computing systems, particularly in environments where processing units differ in computational capabilities. Traditional scheduling methods, such as HEFT \cite{Topcuoglu2002PerformanceEffectiveAL}, prioritize tasks based on estimated execution and communication times, optimizing performance in environments with diverse processing resources. LB-HEFT \cite{Mahmoud2021AnEL} extends this approach by improving load distribution across resources, reducing idle times and enhancing system utilization. Simpler load-balancing techniques, such as Round-Robin and Dynamic Priority-Based Round Robin \cite{Venu2024DynamicPR}, distribute tasks evenly across available devices, preventing overloading of specific processing units. For dependency-constrained systems, like CNNs, longest-path-based scheduling methods, like the Critical Path on a Processor (CPOP) \cite{Topcuoglu2002PerformanceEffectiveAL}, were proposed for minimizing total execution time by prioritizing the most computationally expensive sequence of dependent operations. 

However, in hybrid IMC architectures, the PUs do not usually differ in computational capacity, but rather in functional capabilities. This distinction necessitates a new scheduling approach that optimally assigns tasks/nodes, while ensuring execution balance.  To systematically evaluate CNN task/nodes scheduling across such architectures, we leverage the In-Memory Computing Emulator (IMCE), introduced in \cite{neurosoc_i2mtc_2025}, designed to emulate the behavior of the IMC integrated circuit (IC) developed in the NeuroSoC project \cite{neurosoc}. 



\section{The IMCE Architecture} \label{IMCE_architecture}

\begin{figure}[t!]
\centering
\includegraphics[width=3.2in]{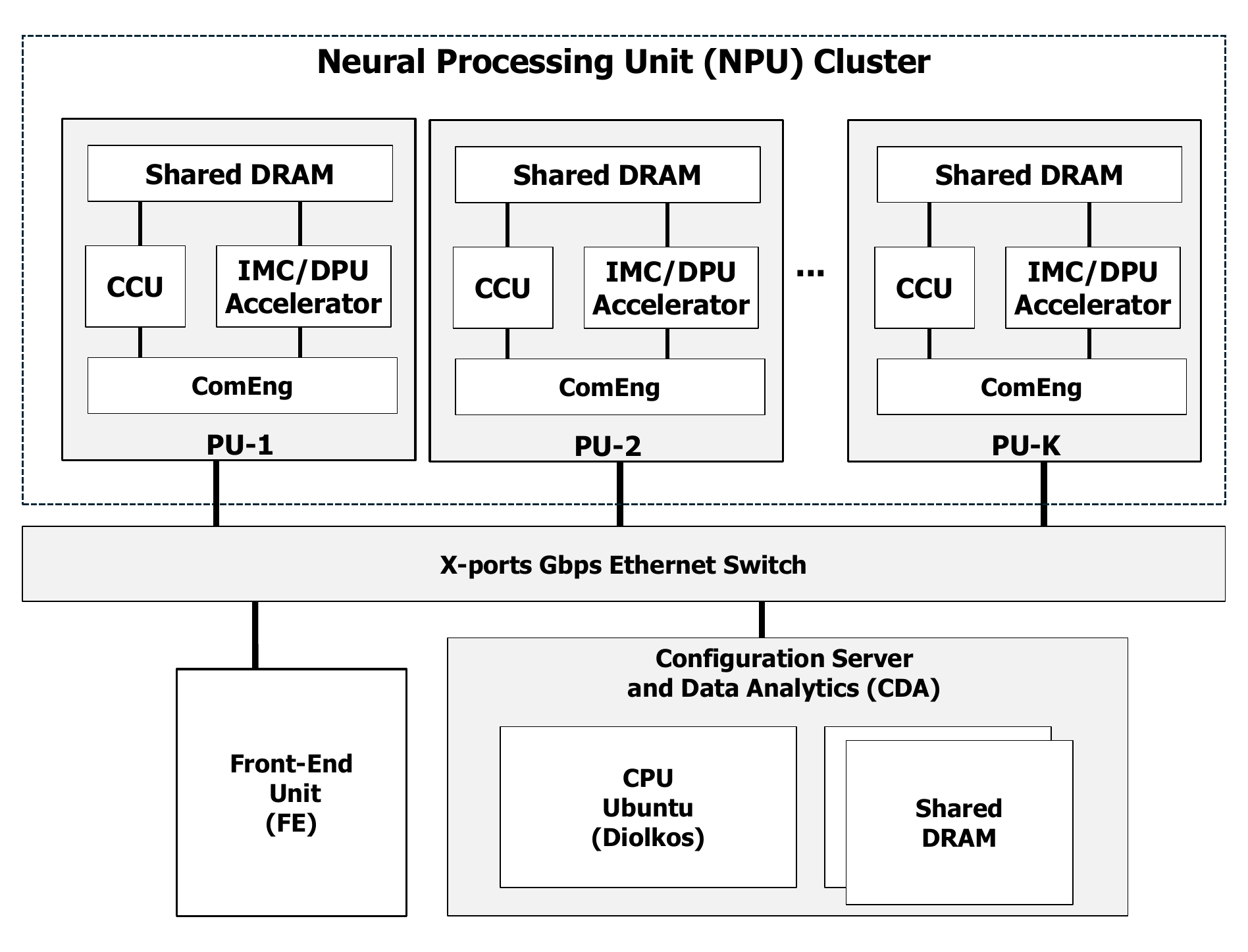}
\caption{The IMCE Architecture} \label{fig:IMCEarch}
\end{figure}

The In-Memory Computing Emulator (IMCE)  is a modular and scalable Deep Neural Network (DNN) emulation platform designed to support a wide range of AI workloads. As shown in Fig.  \ref{fig:IMCEarch},  it consists of three main components: the  Front-End Unit (IMCE-FE), the Processing Units (IMCE-PUs), and the Configuration and Data Analytics Server (IMCE-CDA) \cite{neurosoc_i2mtc_2025}. The IMCE-FE is based on a dedicated FPGA and supports the electrical interfaces of the emulated IC, facilitating communication between the IMCE and the testing platform of the under development IC. 

The IMCE-PUs implement the functions of CNN nodes using either IMC or DPU hardware accelerators. The IMC-PUs handle operations related to matrix-vector multiplications (MVMs) and convolutions (Convs), optionally followed by activation functions such as Rectified Linear (ReLU) or Sigmoid Linear (SiLU). The DPU-PUs are used mainly to execute a rich set of digital operations, like addition, max/average pooling, concatenation, and splitting, although functions similar to IMC-PUs are also supported but with lower performance. These PUs are organized in clusters, where multiple PUs work together to execute DNN workloads efficiently. Several NPU clusters can be further interconnected via high-speed links, forming a large-scale emulation environment that mirrors the architecture of a real in-memory computing chip. The PUs follow a compute-and-forward paradigm, where processing starts as soon as input data arrive, maximizing pipeline efficiency. Each PU runs multiple DNN nodes concurrently using ARM processors, leveraging an inter-processor interrupt (IPI) mechanism for communication and shared DRAM for data exchange. 



The IMCE Configuration Server and Data Analytics (CDA) is a Linux environment that manages data communication via synchronous connections or asynchronous shared memory transactions. The CDA configures the IMCE architecture based on the target DNN, dynamically allocates computational resources, and collects statistical data for performance evaluation and debugging. Detailed description of IMCE is given in \cite{neurosoc_i2mtc_2025}.  

For the deployment of AI models on IMCE, a dedicated software stack was developed to support various AI frameworks and to automate hardware configuration. This stack optimizes, quantizes, and maps trained models. It includes runtime management, ensuring seamless inference execution and system initialization. Additionally, it enables efficient memory access for debugging and statistics collection, while optional noise modeling enhances robustness when AIMCs are emulated.

The IMCE serves as a versatile platform for deploying DNN/CNN models, evaluating inference accuracy, and analyzing various performance metrics. The IMCE also enables the study of diverse aspects, such as task mapping, and computational efficiency, and facilitates extracting valuable insights that contribute to optimized AI model deployment.

\begin{algorithm}[t] 
    \small 
    \caption{Load Balance Longest Path (LBLP)}
    \label{alg:LBLP}
    \begin{algorithmic} 
        \STATE \textbf{Input:} Network nodes, available PUs
        \STATE \textbf{Output:} Node-to-PU mapping

        \STATE \textbf{Step 1: Identify the Longest Path (LP)}
        \STATE The LP is the sequence of nodes that form the path with the highest total execution time.

        \STATE \textbf{Step 2: Sort Nodes by Execution Time and Processing Type}
        \FOR{each Processing Type (IMC/DPU)}
        \STATE Sort the nodes of LP in descending order based on execution time.

        \STATE \textbf{Step 3: Assign LP Nodes to PUs}
            \FOR{each node in the sorted LP list}
                \STATE Assign node to the PU with the smallest total assigned execution time.
                \STATE Update the total assigned execution time of the selected PU.
            \ENDFOR

        \STATE Sort the non-LP nodes in descending order
        \STATE \textbf{Step 4: Repeat the procedure of Step 3 for the non-LP nodes.}
        \STATE Ensure parallel branch constraint is respected.

      \ENDFOR
    \end{algorithmic}
    \normalsize
\end{algorithm}

\section{Nodes Mapping to IMCs} \label{Algorithms}

For  optimizing a CNN execution on hybrid IMC/DPU devices, it is essential to consider how computational workloads are distributed across the available PUs. The most computationally demanding operations—such as Convs and MVMs— are executed on IMCs, while  operations such as additions, pooling, concatenations, and reshaping are mapped to DPU-PUs, since functions are not supported by the IMCs.

For that purpose, we present a new mapping algorithm, called Load Balance Longest Path (LBLP), along with three other algorithms — Weights Balance (WB), Round-Robin (RR), and Random (RD) — which are used to  perform a comparative analysis for different CNNs.  The evaluation of these algorithms was conducted experimentally in the IMCE platform. Each algorithm employs a distinct strategy for assigning the CNN nodes to the available PUs, while adhering to execution constraints.  

The PUs in hybrid architectures are usually of two distinct types, IMCs and DPUs. Each type supports solely a specific set of functions, while functions supported on both types, demonstrate significantly different execution times. 


The Load Balance Longest Path (LBLP) algorithm is based on  node execution times and efficiently assigns the nodes of a network to the available PUs by targeting load balancing. The LBLP steps are presented above in Algorithm \ref{alg:LBLP}. The LBLP identifies the longest path on the network and assigns nodes to PUs using an iterative process, by allocating nodes to PUs that support their specific functionality and result to balanced loading, by using in each step the PU with the minimum total assigned execution time.  
The algorithm also, enforces the constraint that nodes that belong to parallel branches must be assigned, if possible, to different PUs to maximize parallelism.

The Weights Balance (WB) (Algorithm \ref{alg:WB} below) ensures an even distribution of weights (coefficients) among the PUs operating as IMCs and an even distribution of execution times among DPUs.





\begin{algorithm}[h] 
    \small 
    \caption{Weights Balance (WB)}
    \label{alg:WB}
    \begin{algorithmic} 
        \STATE \textbf{Input:} Network nodes, available PUs
        \STATE \textbf{Output:} Node-to-PU mapping

        \STATE \textbf{Step 1: Allocation of IMC Nodes}
            \STATE      Sort the IMC nodes in descending order based on  weights size.
        \FOR{each node in sorted list}
            \STATE Assign node to the PU with the smallest assigned weights size.
            \STATE Update the total weights size of the assigned PU.
        \ENDFOR

        \STATE \textbf{Step 2: Allocation of DPU Nodes}
        \STATE Sort the DPU nodes in descending order based on execution time.
        \FOR{each node in sorted list}
            \STATE Assign node to the PU with the smallest total execution time.
            \STATE Update the total execution time of the assigned PU.
        \ENDFOR

    \end{algorithmic}
    \normalsize
\end{algorithm}

The Round-Robin (RR) algorithm assigns nodes to a set of available PUs in a cyclic manner, ensuring a balanced distribution based on the number of nodes per PU. The algorithm first performs a topological sort on the network to establish a valid execution order and then sorts nodes in ascending order based on their unique node IDs. The nodes are then assigned sequentially to PUs in a round-robin fashion, where the first node is allocated to the first PU, the second node to the second PU, and so on. Once all PUs are occupied, the process cycles back to the first PU and continues until all nodes are assigned. 

The Random (RD) algorithm assigns nodes to PUs in a non-deterministic manner. Initially, a number of nodes equal to the available PUs are randomly selected and assigned to different PUs to ensure full utilization of resources. The remaining nodes are then assigned randomly to a PU. This approach introduces randomness into the mapping process, which can be beneficial in scenarios where workload distribution patterns are highly dynamic and unpredictable. 

Concerning the effect of these algorithms on the total energy consumption, it is estimated that energy consumption is not affected as long as the nodes are executed on the same type of PUs in all allocations.

\section{Performance Results} \label{PerfResults}

For experimenting and validating the effect of various allocation algorithms on the processing performance of CNNs, we used the previously presented IMCE environment. Within the IMCE we can determine the number of available PUs, the number of IMCs and DPUs used and we can set-up the IMCE as an inference engine according to the selected allocation algorithm. We used pre-trained CNN models, quantized to INT8 arithmetic and two type of PUs, IMCs and DPUs. Since the IMCE uses FPGAs for implementing the various PUs, the reconfiguration is achieved by programming the FPGA of each PU with the proper configuration file. Each CNN node is allocated to a specific PU and parameterization of all nodes/PUs is achieved by using custom software tools, that extract the required information from the respective CNN ONNX model, and a generic configuration mechanism.

By varying the number of available PUs, we observe significant differences on the performance metrics, like Normalized Processing Rate and Latency. Normalized Processing Rate is calculated by dividing the measured processing rates by their maximum value, while Normalized Latency is calculated by dividing the measured latencies by their minimum value.

As more PUs are available, the workload is distributed among a larger number of PUs, thus reducing the execution time per PU. However, achieving improved performance requires optimum resources allocation. An inefficient allocation may lead to under-utilized resources or bottlenecks, particularly in CNN models with large convolutions.

To evaluate the performance of the proposed allocation algorithm, we assess three neural network architectures: two residual networks \cite{resnet_cvpr_2016}, ResNet8 and ResNet18, and one convolutional neural network \cite{ultralytics}, YOLOv8n, as described below.

\subsection{ResNet8}
ResNet8 is the smallest variant of the ResNet family, containing 14 nodes in total, 10 of which are convolutional. This network has 78K parameters (including weights and biases). For classification, we use the CIFAR-10 dataset, consisting of 32x32 pixel images spanning 10 distinct classes. 

\begin{figure}[hbt!]
\centering
\includegraphics[width=3.25in]{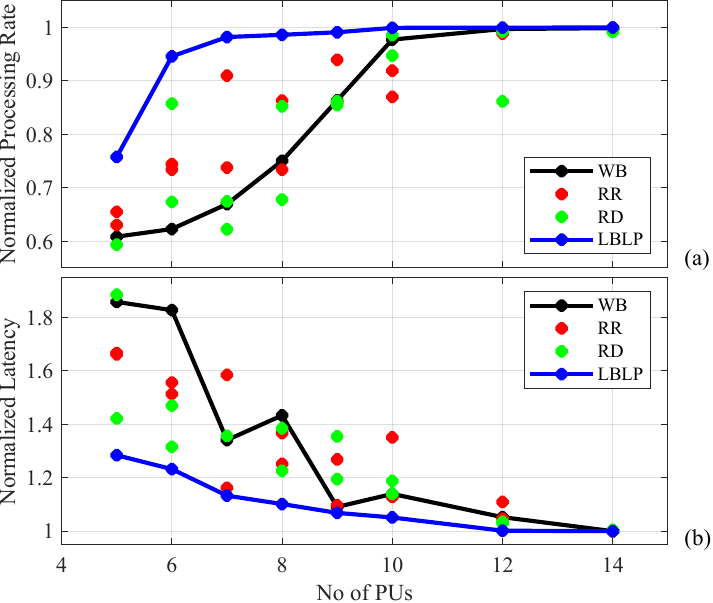} 
\caption{ResNet8: Normalized Processing Rate (a) and Latency (b) vs Number of PUs for various allocation methods} \label{fig:Resnet8_rate}
\end{figure}

Figure \ref{fig:Resnet8_rate} presents the Normalized Processing Rate and Latency of ResNet8 versus the number of available PUs for all four scheduling algorithms: LBLP, WB, RR, and RD. As the number of available PUs increases, fewer nodes of the network are assigned to each PU, reducing the total execution time per PU.  As a result, the processing rate and latency of all algorithms converge to the same value when the number of PUs equals 14, where each PU processes exactly one node.

Among all algorithms, LBLP consistently delivers the best performance in both metrics. Notably, LBLP achieves the highest processing rates across all configurations. Additionally, LBLP maintains the lowest latency across all configurations, outperforming WB in both metrics and demonstrating superior overall efficiency. Meanwhile, RR and RD exhibit moderate performance, occasionally surpassing WB in certain configurations but consistently falling short of LBLP’s performance across all cases.


\subsection{ResNet18}
ResNet18 is another, more complex variant of Residual Networks, compared to ResNet8. For consistency in our evaluation, we use again the CIFAR-10 dataset instead of ImageNet, which typically processes images of 224x224 pixels.To accommodate the smaller image size, model parameters have been adjusted accordingly. In that case, ResNet18 consists of 30 nodes, with 20 convolutional layers (including 11 with ReLU activations), and a total of 2.8M parameters.

\begin{figure}[t!]
\centering
\includegraphics[width=3.2in]{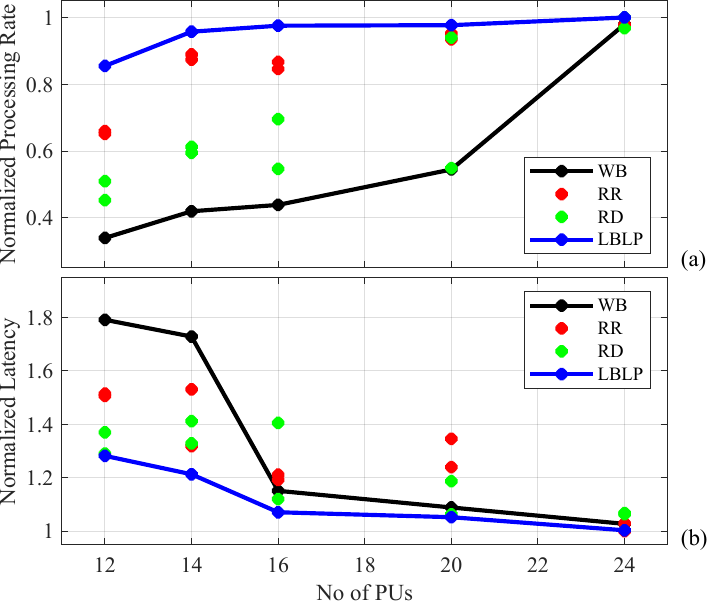} 
\caption{ResNet18: Normalized Processing Rate (a) and Latency (b) vs Number of PUs for various allocation methods.} \label{fig:Resnet18_rate}
\end{figure}

As shown in Figure \ref{fig:Resnet18_rate},  the four scheduling algorithms exhibit similar trends  to those observed in ResNet8. LBLP consistently achieves the highest performance in both processing rate and latency, demonstrating its efficiency in distributing computational workloads effectively.

In Table \ref{tab:table_1} it is shown how the CNN nodes have been allocated to the different PUs, when LBLP and WB are used, for an IMCE with 12 PUs in total. In this configuration 8 PUs are used for convolutions (20) and MVM (1). This Table also shows how the weights of the corresponding nodes are distributed to the different PUs and what is the achieved utilization  per PU.  According to Figure \ref{fig:Resnet18_rate}, LBLP achieves more than 2x processing rate and x1.4 less latency, compared to WB. This is achieved due to better and more uniform utilization of all PUs  (78.3\% mean utilization for LBLP and 24.4\% for WB).

\begin{table*}[t!]
    \centering
    \caption{ResNet18: The distribution of MVM nodes to PUs, normalized weights area and normalized utilization}   \label{tab:table_1}
    \begin{tabular}{|l r|l|l|l|l|l|l|l|l|}
        \cline{1-10}
        Proc. Unit & & 1 & 2 & 3 & 4 & 5 & 6 & 7 & 8 \\
        \hline
        LBLP & Nodes & 6, 15, 22 & 3, 8 & 5, 20 & 2, 19 & 9, 16, 24 & 1, 13, 26 & 10, 27, 30 & 12, 17, 23  \\
          & Weights Area [\%] & 7.4 & 1.6 & 23.0 & 23.0 & 100.0 & 92.0 & 92.3 & 70.3  \\
          & Utilization [\%] & 91.6 & 88.8 & 99.5 & 100.0 & 81.6 & 83.2 & 72.4 & 99.4  \\
        \hline{}
        WB  & Nodes & 24 & 26 & 27 & 23 & 13, 17, 30 & 15, 19, 22 & 1, 3, 6, 8, 9, 20 & 2, 5, 10, 12, 16  \\
         & Weights Area [\%] & 100.0 & 100.0 & 100.0 & 50.0 & 31.7 & 31.9 & 31.7 & 28.1  \\
          & Utilization [\%] & 17.4 & 11.1 & 12.4 & 15.3 & 25.2 & 13.1 & 91.0 & 100.0\\
        \hline
    \end{tabular}
\end{table*}

\begin{figure}[t!]
\centering
\includegraphics[width=3.2in]{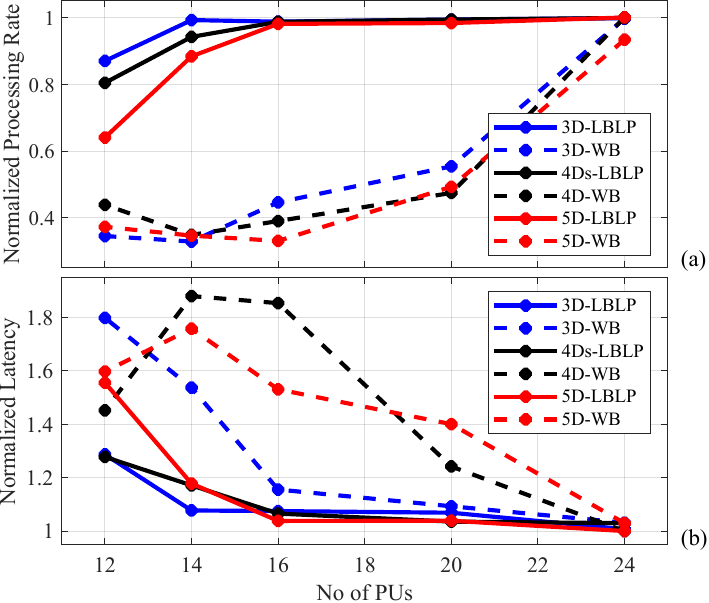} 
\caption{ResNet18: Normalized Processing rate (a) and latency (b) vs Number of PUs  for different number of DPUs} \label{fig:Resnet18_dpus}
\end{figure}

During the design of an edge AI chip with multiple PUs, and for a given footprint of the available IMC and DPU cores, a question that usually arises is how the available chip area will be utilized, namely how many IMC and DPU cores will be used and how this will affect the chip's performance when different NNs will be executed. Figure \ref{fig:Resnet18_dpus} shows the rate and latency metrics for different combinations of IMCs and DPUs, in the IMCE experimental environment. As the number of DPUs increases, the number of IMCs, available for MVM and convolutions, decreases for a given total number of PUs. That scenario represents the case of accelerators with fixed number and types of PUs. In all cases LBLP shows significantly improved performance compared to WB.



\subsection{YOLOv8n}
YOLOv8n is the smallest variant of the YOLO  models, one of the best performing object detectors, optimized for speed and efficiency in IoT applications. In this work, we analyzed a subset of this network, that consists of 233 nodes, with 63 of them convolutional  (57 followed by SiLU activations), with a total of 3.17M parameters. The YOLOv8n model is mostly sequential and only partially contains parallel branches. Specifically, it features 3 parallel main branches, each containing two short sub-branches  and one long sub-branch.  Each short sub-branch consists of three convolutional nodes and each long sub-branch consists of five convolutional nodes, so parallelism may affect at most 10\% of the total latency. By experimenting with the LBLP and WB strategies, up to 6\% difference in latency has been measured.

\section{Conclusions} \label{Conclusions} 
The performance of CNN inference engines is determined not only by their raw processing capabilities, but also by the resources utilization, namely on how a CNN model is mapped to the internal architecture, especially when multiple processing units can operate in parallel. In this work we presented a resources optimization algorithm and using some CNNs, we demonstrated the improvement achieved on processing rate and latency. The proposed algorithm is of low complexity, it exploits the inherent parallelism of the CNN graph, and based on measured execution times, it balances the resources utilization and maximizes parallelism during inference. The experimental results were based on an accurate emulation setup and further experimentation on a hybrid integrated circuit are planned when the chip will be available.

\section{Acknowledgments}
This work has been performed in the framework of the EU project "NeuroSoC: A multiprocessor system on chip with in-memory neural processing unit", HORIZON-101070634 \cite{neurosoc}.

\bibliographystyle{IEEEtran}
\bibliography{mocast_2025}

\end{document}